          [2001/04/25 1.1 (PWD)]
\documentclass[a4paper,twocolumn,dvips]{esapub} 
\usepackage{natbib}
\usepackage{epsfig}
\usepackage{latexsym}
\usepackage{graphicx}
\title{SPI energy calibration}
\author{V. Lonjou}
\author{J. Kn{\"o}dlseder}
\author{J.P. Roques}
\author{G.K. Skinner}
\author{P. von Ballmoos}
\author{P. Jean}
\author{P. Paul}
\author{G. Weidenspointner\footnote{ESA Fellow}}
\affil{Centre d'{\'E}tude Spatiale des Rayonnements, BP 4346, 31028
Toulouse, France} 
\author{C. Wunderer}
\affil{University of California, Berkeley, CA 94720, USA}
\author{S. Schanne}
\affil{CEA Saclay, DSM/DAPNIA/Service d'Astrophysique, 91191 Gif-sur-Yvette, France}

\begin{document}

\keywords{SPI; energy calibration; energy; space between}
\maketitle

\begin{abstract}
The status of the SPI energy calibration after the first year of INTEGRAL operations is reported. We have studied the gain variations and we have demonstrated that the most important parameter is the germanium detector temperature. This study permits us to determine the limits of our calibration method and the frequency of calibrations needed.
\end{abstract}

\section{Introduction}
\label{intro}
The Spectrometer for INTEGRAL (SPI) is one of the two main instruments on board the ESA INTEGRAL mission. A detailed description of SPI can be found in \citet{vedrenne03}. SPI is a coded-mask telescope with a fully-coded field of view of $16^\circ$ that uses a 19 pixels high-resolution Ge detector. The detectors cover the energy range 20-8000 keV with an energy resolution of 2-8 keV FWHM. We report here the study of the variations of the SPI energy calibration, and the status of the calibration method.

\section{Temperature and time dependence of the gain}
\label{tt_gain}

It is well known that the temperature of the germanium crystals has an influence on their forbidden band. The band gap of germanium crystal changes with temperature, thus altering the charge released by a given energy deposit, and hence the calibration.

\begin{figure}[!t]
  \center
  \epsfxsize=7cm \epsfclipon
  \epsfbox{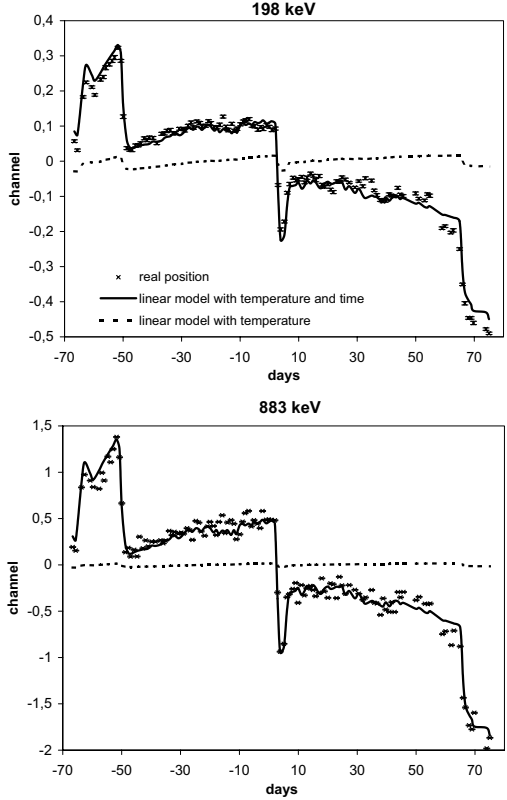}
  \caption{\label{fig:pos_198_883}
  Position of the 198 keV and the 883 keV lines as a function of time. The origin of time is the middle of the revolution 60.
  }
\end{figure}

\begin{figure}[!t]
  \center
  \epsfxsize=7cm \epsfclipon
  \epsfbox{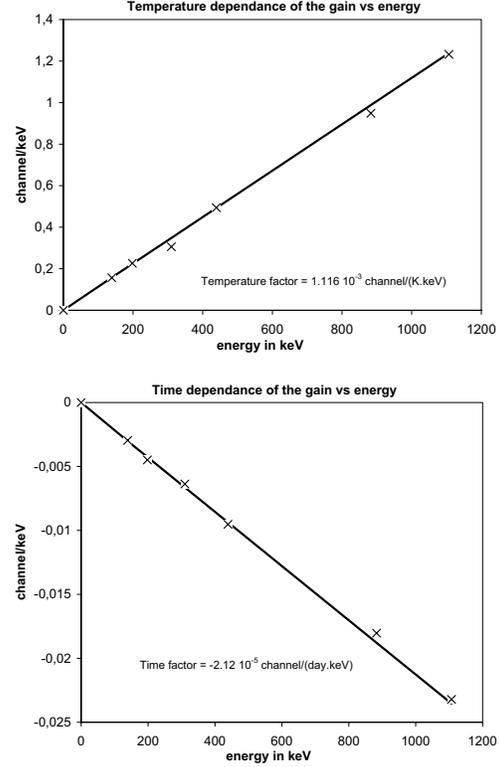}
  \caption{\label{fig:factor}
  Temperature and time dependence of the calibration. We also show  the best fit linear function. The slopes give the temperature-factor and the time-factor.
  }
\end{figure}

A preliminary study of raw spectra in ADC channels ($\simeq$0.135 keV) was made in order to determine the relation between the position of background lines and the temperature of the germanium detectors.
The choice of the lines was based on the SPI background line identifications \citep{weidenspointner03} and instrumental background simulations employing the MGGPOD Monte Carlo suite \citep{weidenspointner04a, weidenspointner04b} to identify clean lines.
It is apparent from Figure~\ref{fig:pos_198_883} that a model having a simple linear dependence on temperature cannot fit the data.
Neither a second nor a third order polynomial was satisfactory. However, a model in which the line shift is a combination of a linear function of temperature and a linear function of time works very well as shown in Figure \ref{fig:pos_198_883}. More complex time and temperature dependences have been tested but do not improve the fit significantly. Each line study provides 2 parameters: one for the temperature and one for the time.
The linear dependence of these two parameters (Figure~\ref{fig:factor}) on energy indicates that the calibration variations can be considered as a gain variations. Thus two factors are sufficient to explain the gain fluctuations: the temperature-factor $C_{Temp}$=$1.116 \times 10^{-3}$~channel~K$^{-1}$~keV$^{-1}$ and the time-factor $C_{Time}$=$-2.12 \times 10^{-5}$~channel~day$^{-1}$~keV$^{-1}$.

The proposed deterministic calibration model has been tested on the data between revolutions 43-90 and on the data between revolutions 96-110. In each case, one revolution was taken as a reference and corrections based on temperature and time were applied to obtain calibration factors for the other revolutions. We obtain an accuracy which is always better than 0.35 keV (see Table 1) for one detector; if the data are summed over all the detectors the errors are reduced to less than 0.1 keV. The errors are dominated by statistical fluctuations in the calibration for the reference revolution. Thus the apparent improvement when averaging over all the detectors is not surprising.
\begin{table*}
  \begin{center}
    \renewcommand{\arraystretch}{1.2}
    \begin{tabular}[h]{lcccc}
      \hline
      line in keV 				 	& 198 	& 883  & 1764 	& 2754 \\
      \hline
      Absolute accuracy in keV for one detector		& 0.05	& 0.2  & 0.3	& 0.35 \\
      Absolute accuracy in keV for the detector plane	& 0.05	& 0.05 & 0.1	& 0.1 \\
      \hline \\
      \end{tabular}
    \label{tab:acc_det}
    \caption{Maximum value of the calibration errors when using a deterministic calibration }\vspace{1em}
  \end{center}
\end{table*}

Detector degradation changes the shape of the line and therefore has an influence on the energy calibration. We have investigated the magnitude of this effect by comparing the line energy deduced by fitting a Gaussian function either to a narrow region around the maximum or to a large band around the line. The results are shown in Table~\ref{tab:error_deg}, one can note that the magnitude of this effect is dependent on the line model. The changes of the line shape due to the detector degradation are not sufficient to fully explain the time-factor but constitute a significant part of it.
\begin{table}
  \begin{center}
    \renewcommand{\arraystretch}{1.2}
    \begin{tabular}[h]{lccc}
      \hline
      energy in keV		&  883  & 1764 	& 2754 \\
      \hline
      Gaussian + constant 	& 0.03  & 0.1	& 0.15 \\
      Gaussian + linear function&$\ge$ 0.2  & $\ge$ 0.2& $\ge$ 0.2 \\
      \hline \\
      \end{tabular}
    \label{tab:error_deg}
    \caption{1 $\sigma$ error on the centroid in keV as a function of energy and of line model}\vspace{1em}
  \end{center}
\end{table}

We conclude that the variations of the SPI energy calibration are well modelled. These variations are equivalent to a gain drift and they have two different origins, temperature and time. The effect of the variation of the forbidden band in the Germanium is of the same order of magnitude as the temperature effect and the «time-factor»  is partly explained by the degradation.

\section{Status of the calibration method}
\label{status}

\begin{figure}[!t]
  \center
  \epsfxsize=7cm \epsfclipon
  \epsfysize=3.5cm \epsfclipon
  \epsfbox{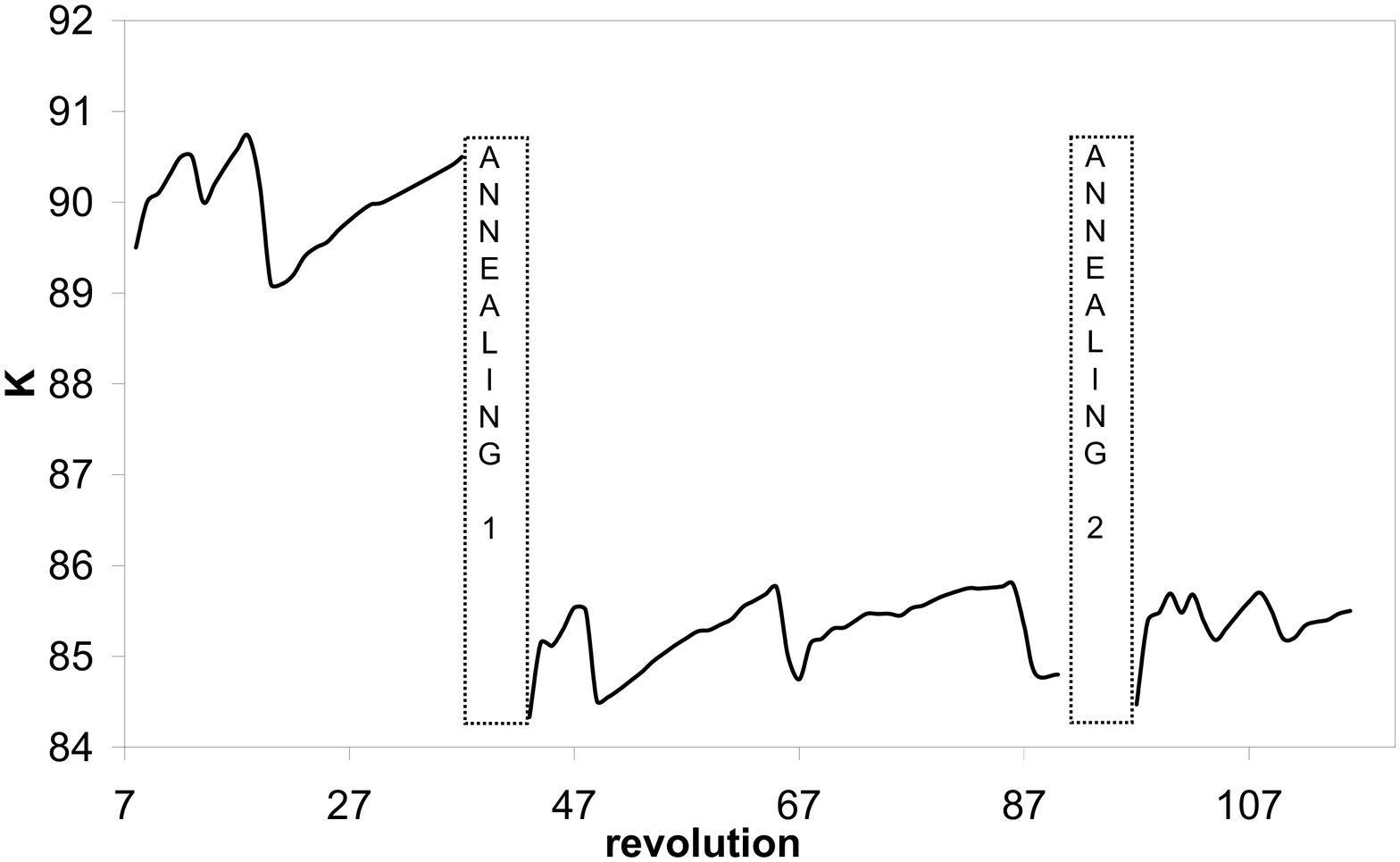}
  \caption{\label{fig:tcp_par_orbite}
  Temperature of the germanium detector as a function of revolution number.
  }
\end{figure}

\begin{figure}[!t]
  \center
  \epsfxsize=7cm \epsfclipon
  \epsfysize=3.5cm \epsfclipon
  \epsfbox{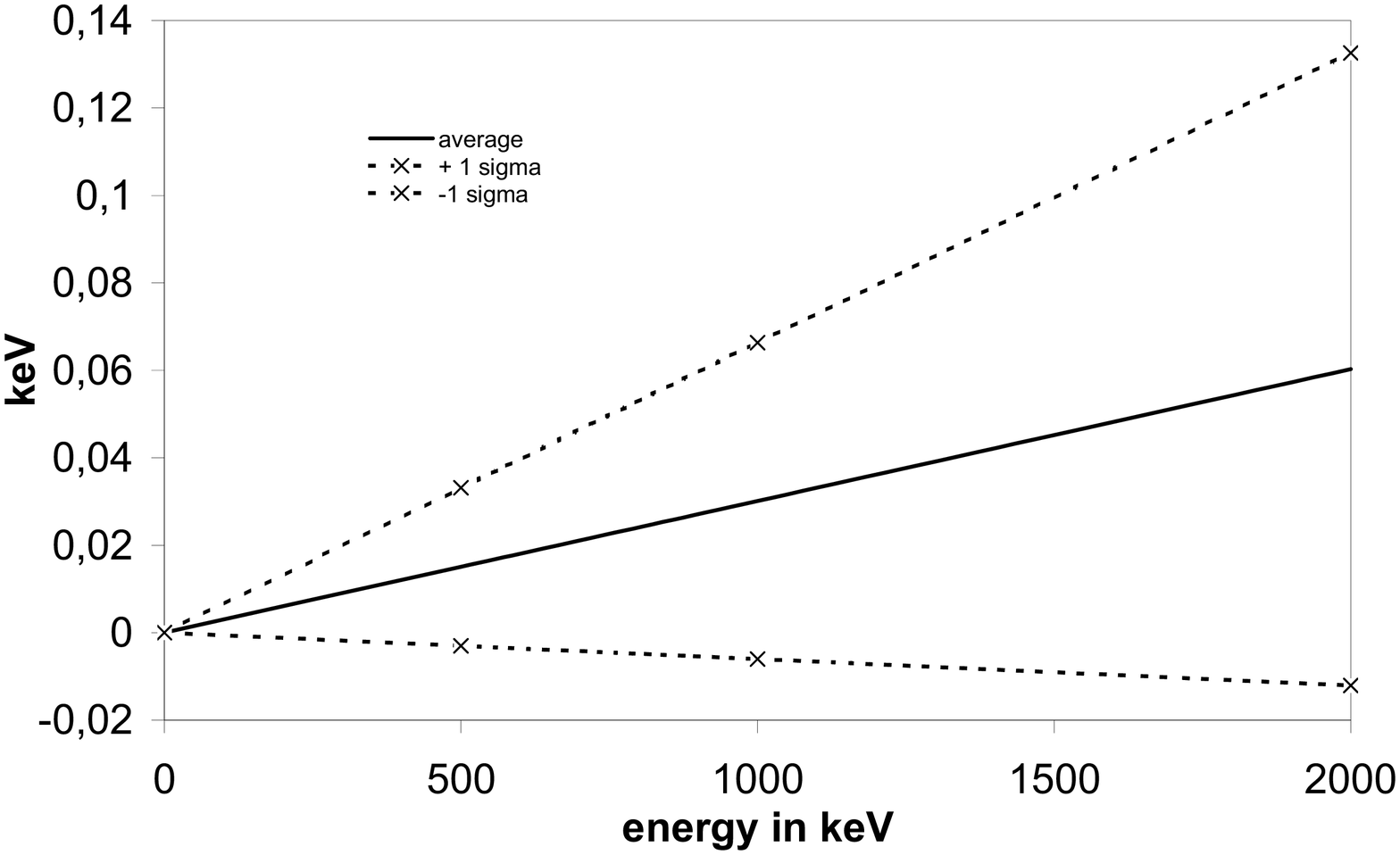}
  \caption{\label{fig:typ_gain_temp_1orbit}
  Shift due to temperature variations during a typical orbit. This graph is obtained by multiplying the variation of temperature by $C_{Temp}$.
  }
\end{figure}

\begin{figure}[!t]
  \center
  \epsfxsize=7cm \epsfclipon
  \epsfysize=3.5cm \epsfclipon
  \epsfbox{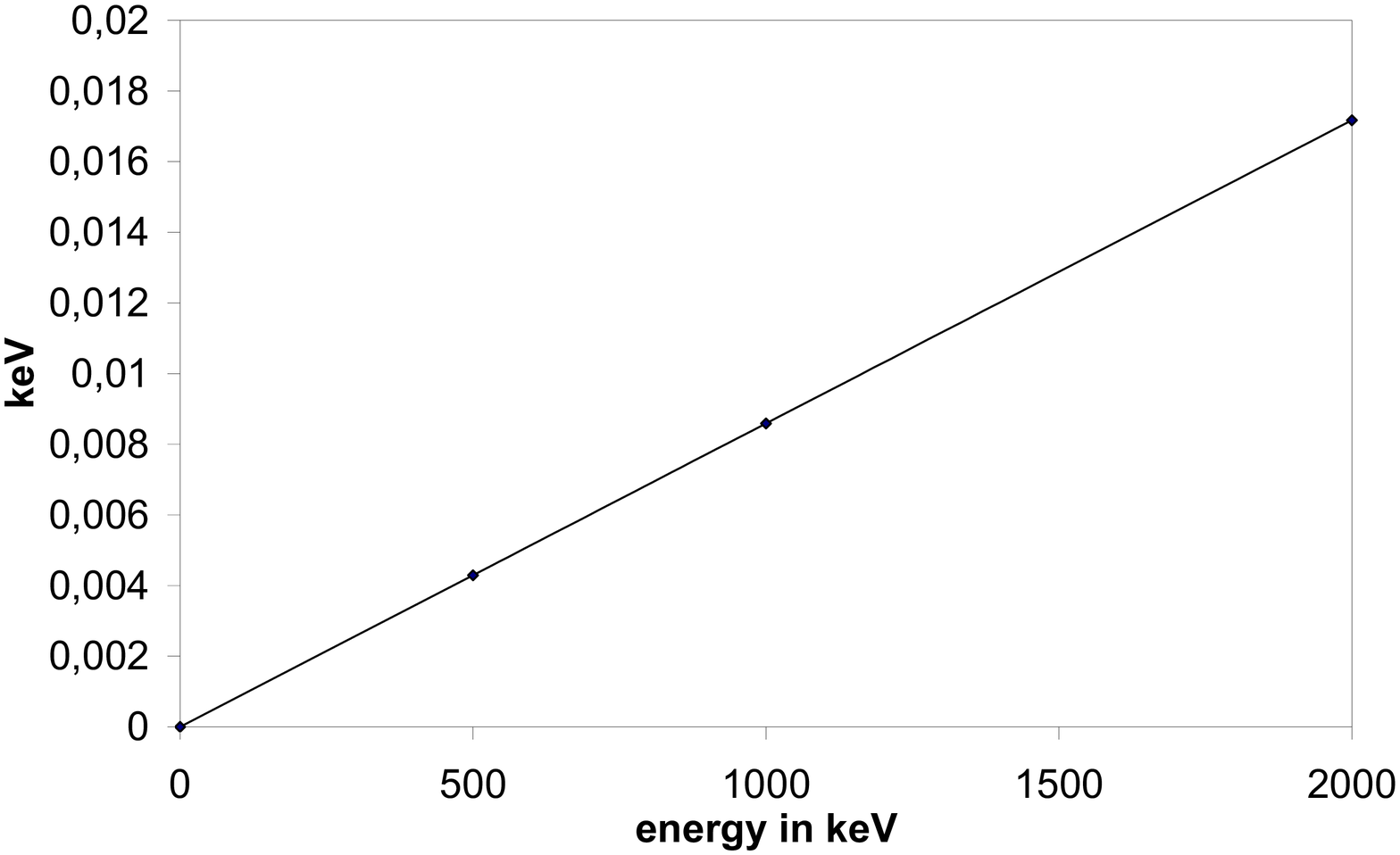}
  \caption{\label{fig:typ_gain_time_1orbit}
  Shift due to $C_{time}$ during a typical orbit.
  }
\end{figure}

The above analysis permits us to define the limits of our calibration method and the frequency of calibrations needed. The primary effect is the temperature. Figure~\ref{fig:tcp_par_orbite} shows the variations of the detector temperature as a function of time. One can see that the temperature variations on time scales of the order of one revolution are usually relatively small. The mean value of the absolute difference between the maximum and minimum temperatures within one revolution is 0.20 K. Figure \ref{fig:typ_gain_temp_1orbit} shows the effect of such a variation as a function of energy. The lines corresponding to values of temperature difference which are larger or smaller than this mean by 1 $\sigma$ are also shown. 
We conclude that the error due to temperature changes within one orbit is $\le$0.15 keV in the range 20 keV -- 3 MeV. The shift in the calibration due to the time-factor within one orbit is always negligible ($\simeq$0.01 keV at 1 MeV, see Figure \ref{fig:typ_gain_time_1orbit}). 

\begin{figure}
	\includegraphics[height=3.5cm,width=8cm]{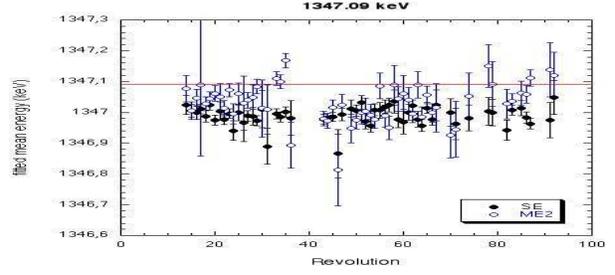}
	\caption{Position of the 1347 keV line as a function of revolution number for single and double events. The theoretical position is shown for comparison}
	\label{fig:1347}
\end{figure}

\begin{figure}
	\includegraphics[height=3.5cm,width=8cm]{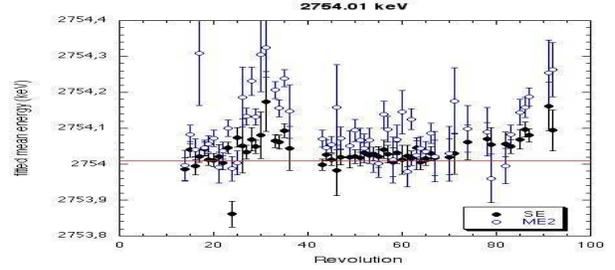}
	\caption{Position of the 2754 keV line as a function of the revolution number for single and double events. The theoretical position is shown for comparison}
	\label{fig:2754}
\end{figure}

We conclude that one solution  per orbit and per detector is sufficient. The calibration has been checked by using so-called multiple events. These are events in which energy is deposited in more than one detector. Looking at the calibration for the multiple events for a single high-energy line provides a check on the calibration over the entire energy range of SPI up to that line energy. We conclude (see Figure \ref{fig:1347} and Figure \ref{fig:2754}) that the absolute calibration is better than 0.15 keV over all the energy range  below 2.7 MeV for a spectrum summed over one orbit.


\section{Conclusion}
\label{conclusion}
The variation of the SPI energy calibration is dominated by two factors: the temperature of the germanium detector and a time-factor. One calibration per orbit and per detector are sufficient to yield an accuracy better than 0.15 keV. We are now going to further improve the calibration method by implementing software which takes into account both effects on time scales shorter than one orbit.

\section*{Acknowledgments}

Based on observations with
INTEGRAL, an ESA project with instruments and science data centre
funded by ESA member states (especially the PI countries: Denmark,
France, Germany, Italy, Switzerland, Spain), Czech Republic and
Poland, and with the participation of Russia and the USA.


\bibliographystyle{aa}
\bibliography{esapub}

\end{document}